\documentclass{article}

\usepackage{PRIMEarxiv}

\usepackage[utf8]{inputenc} 
\usepackage[T1]{fontenc}    
\usepackage{hyperref}       
\usepackage{url}            
\usepackage{booktabs}       
\usepackage{amsfonts}       
\usepackage{nicefrac}       
\usepackage{microtype}      
\usepackage{lipsum}
\usepackage{fancyhdr}       
\usepackage{graphicx}       
\graphicspath{{media/}}     
\usepackage{listings}
\usepackage{tcolorbox}
\title{FALQU: Finding Answers to Legal Questions}
\author{
Behrooz Mansouri\\
University of Southern Maine\\
Portland, Maine, USA\\
\texttt{behrooz.mansouri@maine.edu}
\And
Ricardo Campos\\
Polytechnic Institute of Tomar \\
Ci2 - Smart Cities Research Center / INESC TEC\\
Tomar, Portugal\\
\texttt{ricardo.campos@ipt.pt}
}

\begin{document}
\maketitle

\begin{abstract}
This paper presents a new test collection for Legal IR, FALQU: \textbf{F}inding \textbf{A}nswers to \textbf{L}egal \textbf{Q}uestions, where questions and answers were obtained from Law Stack Exchange (LawSE), a Q\&A website for legal professionals, and others with experience in law. Much in line with Stack overflow, Law Stack Exchange has a variety of questions on different topics such as copyright, intellectual property, and criminal laws, making it an interesting source for dataset construction. Questions are also not limited to one country. Often, users of different nationalities may ask questions about laws in different countries and expertise. Therefore, questions in FALQU represent real-world users' information needs thus helping to avoid lab-generated questions. Answers on the other side are given by experts in the field. FALQU is the first test collection, to the best of our knowledge, to use LawSE, considering more diverse questions than the questions from the standard legal bar and judicial exams. It contains 9880 questions and 34,145 answers to legal questions. Alongside our new test collection, we provide different baseline systems that include traditional information retrieval models such as TF-IDF and BM25, and deep neural network search models. The results obtained from the BM25 model achieved the highest effectiveness.

\end{abstract}

\section{FALQU Test Collection}
Despite being a recent research area, legal information retrieval has been at the forefront of research efforts with the surge of a few question-answering legal datasets. The most notable, are COLIEE-2015 \cite{kim2015coliee}, which uses Japanese Legal Bar exams and JEC-QA \cite{zhong2020jec}, which uses questions from the National Judicial Examination of China. Notwithstanding the emergence of these initiatives, datasets still lack diversity in terms of the questions posed and the domains addressed. Platforms such as Stack Exchange have proved to be a good solution to this problem by providing community question-answering networks for different domains. Such networks have been used over the years within the context of Code Summarization \cite{iyer2016summarizing} and Math Information Retrieval \cite{mansouri2022overview} tasks. However, despite their usefulness, community question-answering websites have never been used for legal information retrieval purposes. 

In this work, Law Stack Exchange\footnote{\url{https://law.stackexchange.com/}} (LawSE) is used to build a new test collection for legal information retrieval, a task that is generally understood as the process of finding answers to legal questions or a single answer as is the case in LawSE. This differs from common IR tasks, where the user is usually interested in retrieving the most relevant documents and not a particular one. Such a task can be formally defined as follows:  given a legal question, represented by the question's title and body, an IR model should be able to find (search for) and retrieve the relevant answer (qualified as such by the asker) among all the answers (posts) available in the reference dataset. To build our test collection, we used the 08-Oct-2022 snapshot of LawSE obtained from the Internet Archive. \footnote{\url{https://archive.org} The referred collection is, due to Internet Archive policies, granted for scholarship and research purposes.}
Such snapshot contains 24,187 law questions, with 10,129 having an accepted answer (qualified by the asker as a relevant one). As a means to eliminate duplicate questions from the dataset, we resorted to the available duplicate links feature. Duplicate links refer to links that point to the same (or almost similar) question that has already been posted. After this curated process, we end up with a collection made of 9,880 questions with an accepted answer.

To select the questions for the training and test set, we first split the total set of 9880 questions into 10 bins based on the questions' scores.\footnote{This score, which is a feature of LawSE, is computed as the difference between all the positive and negative votes given by all the users (not specifically the asker), ranging from -9 to 226 in this snapshot.} 

Binning by score can guarantee that questions of training and test set contain questions of similar quality. After binning, from each bin, we randomly (with a fixed seed to guarantee reproducibility) split 90\% of questions for the training set (8892 questions) and the remaining 10\% for the test set (988 questions). Each set has a TREC-formatted QREL file in a Tab Separated Value (TSV) file with four columns: query-number 0 document-id relevance-score; where query-number is the question id, document-id is the answer id, and relevance-score is always 1. In our setting, there is only one answer in th QREL for each question. Such answer is considered the relevant one (with a relevance score of 1) as it is the accepted answer qualified as such by the asker. Any other answer not found in the QREL file can be considered non-relevant.

After generating the training and test questions, we then compiled all the answers (among all the posts obtained for the 9880 questions), resulting in 34,145 answers. 

The compiled answers are provided in TREC format, with tags <DOC>, <DOCNO> which is the actual LawSE answer (post) id, and <TEXT>. Questions are provided in XML format with each question having the <ID> tag that is the actual post id on LawSE, plus the <TITLE>, and the <BODY> tags having the actual LawSE title and body of the question with its corresponding text. Figure \ref{fig:collection} shows a sample question (upper part of the figure) and a sample answer (bottom) in the FALQU test collection. Both FALQU test collection as well as the code to generate it have been made publicly available on GitHub for research purposes.\footnote{\url{https://github.com/AIIRLab/FALQU}} For ease of use, we have separated the training and test topics files along with their related QRELs.

\begin{figure}[!tb]
\centering
\begin{tcolorbox}[width=120mm,height=33mm, top=0pt,left=2pt]
\begin{lstlisting}
...
<Question>
  <ID>17243</ID>
  <TITLE>Should a lease letter ...</TITLE>
  <BODY>I am signing a lease ...</BODY>
</Question>
...
\end{lstlisting}
\end{tcolorbox}
\centering
\begin{tcolorbox}[width=120mm, height=30mm, top=0pt,left=2pt]
\begin{lstlisting}
...
<DOC>
  <DOCNO> 12 </DOCNO>
  <TEXT> Internationally, according to... </TEXT>
</DOC>
...
\end{lstlisting}
\end{tcolorbox}
\vspace{-3mm}
\caption{Example question and answer in FALQU test collection files.
}
\label{fig:collection}
\end{figure}
 
A brief analysis of the dataset, shows that FALQU questions have a variety of subjects, from simple questions such as ``If a malicious website steals my credit card info, what happens?'' to more complex ones involving reasoning and historical knowledge, such as ``How would the actions of H{\"a}nsel and Gretel in the Grimm tale be interpreted in modern law?''. There are also questions specific to the law of a specific country, such as ``As an Iranian, can I sign an Independent contractor agreement, and work remotely for a EU company from Iran?''. Note that all the questions and answers are in English.
\begin{table}
\vspace{-4mm}
  \caption{Mean Reciprocal Rank(MRR) @1000 and Precision@1 for baselines models on test questions.}

  \label{tab:results}
 \centering
  \begin{tabular}{l|c|c}
    \toprule
    Model& MRR@1000 & P@1 \\
    \midrule
    TF-IDF & 0.352 & 0.274\\
    BM25 & 0.349&  0.270\\
    \midrule
    TF-IDF (YAKE) &  0.407& 0.313\\
    BM25 (YAKE) & \textbf{0.414}&  \textbf{0.323}\\
    \midrule
    distilroberta&0.337& 0.243\\
    all-MiniLM-L12-v2&0.372&0.293\\
    \midrule
    distilroberta (Fine-tuned)& 0.368& 0.283\\
    all-MiniLM-L12-v2  (Fine-tuned)& 0.363 & 0.276\\
    \bottomrule
\end{tabular}
\end{table}

\section{Baseline Models}
We provide several baseline systems, including traditional IR models such as BM25 and TF-IDF and two BERT-based
models, using Sentence-BERT \cite{reimers2019sentence}. For traditional retrieval models, we consider 2 types of queries: (1) Question titles as the query, and (2) Keywords extracted from the question body as a bag of words + the question title.

To extract keywords, we used YAKE \cite{campos2020yake} keyword extraction algorithm. Top-5 keywords were extracted per question.  Then, to compute the similarity between questions and answers, we resort to TF.IDF and BM25 PyTerrier \cite{pyterrier2020ictir} implementation models.
For Sentence-BERT, we considered two pre-trained models, `all-distilroberta-v1' and `all-MiniLM-L12-v2'. We fine-tuned both models using all the questions in the training set. For each question, the positive pair is the question and its accepted answer, and the negative pair is the question and a random answer (any other answer than the accepted answer) in the collection. We used 100 epochs and split the training data into 90-10 percent training and validation sets. The best parameters minimize the combination of two loss functions, contrastive \cite{liu2021learning} and multiple negatives ranking \cite{henderson2017efficient} loss.

The systems' effectiveness is compared using two measures: Mean Reciprocal Rank (MRR@1000) and P@1. We choose top-1000 per TREC run criteria of retrieving top-1000. These two measures fit the purposes of this task as there is only one relevant answer per question. The final results are shown in Table \ref{tab:results} using macro-average values. As shown, the highest MRR@1000 (per TREC tasks standard) and P@1 are achieved using the BM25 model with YAKE, significantly better than all the other baselines, except for TF-IDF with YAKE, using the student t-test with p-value$<0.05$. Looking at BERT-based models, fine-tuning the models could provide a slight improvement for the `distilroberta' model. Still, both fine-tuned and pre-trained Bert-based models are less effective when compared to the traditional IR models. Table \ref{tab:samples} shows two questions for which BM25+YAKE retrieved relevant and non-relevant answers. When looking at the instances where P@1=0, one can observe that baseline models were able to retrieve answers that might be relevant, but they are not considered as the accepted answer or were answers given to other similar questions. This yields the importance of manual annotation of candidate answers, which will be left for future work.

\begin{table*}  
  \caption{Sample Questions with P@1=1 (+Answer) and P@1=0 (-Answer) with BM25 with YAKE. (Questions' titles shown)}
  \label{tab:samples}
 \resizebox{\columnwidth}{!}{
 \centering
  \begin{tabular}{l|l}
    \toprule
    Question &  Child Arrangements Order non-biological relative living arrangements \\
    \midrule
     + Answer  & This means that you have the right to make arrangements to do things like arrange for the child to travel...\\
    \midrule
    \midrule
     Question &  Notice period after tenancy agreement runs out \\
    \midrule
     - Answer  & With respect to disciplining its students and employees, a private school can basically do whatever it wants...\\
    \bottomrule
\end{tabular}
}
\end{table*}

\section{Conclusion}
In this paper, we introduced and made available FALQU (Finding Answers to Legal Questions) a new test collection, which contains 8892 training and 988 test questions along with a relevant answer for each question. Further to this, we have conducted an experiment with different baselines including TF-IDF, BM25, and Sentence-BERT models for this task. To measure effectiveness, we considered two measures: Precision@1 and Mean Reciprocal Rank@1000. BM25 model with question title and keywords from the question body achieved the highest effectiveness, considering both measures. We hope FALQU can be used in the future by researchers in the legal information retrieval field and extend this test collection for further usage than retrieval, such as legal question answering where an answer is generated rather than being retrieved.

\section*{Acknowledgments}
Ricardo Campos was financed by National Funds through the FCT - Fundação para a Ciência e a Tecnologia, I.P. (Portuguese Foundation for Science and Technology) within the project StorySense, with reference 2022.09312.PTDC.

\bibliographystyle{unsrt}  
\bibliography{references}

\end{document}